\documentstyle[prb,aps,epsf,psfig,multicol]{revtex}

\begin{document}
\draft

\title{Superconductivity in the SU(N)  
Anderson Lattice at $U=\infty$}

\author{M. A. N. Ara\'ujo}
\address{Departamento de F\'{\i}sica, Universidade de \'Evora,
Rua Rom\~ao Ramalho, 59, P-7001 \'Evora Codex, Portugal}

\author{N. M. R. Peres}
\address{Departamento de F\'{\i}sica, Universidade de \'Evora,
Rua Rom\~ao Ramalho, 59, P-7001 \'Evora Codex, Portugal}

\author{P. D. Sacramento}
\address{Departamento de F\'{\i}sica and  CFIF, Instituto Superior 
T\'ecnico, Av. Rovisco Pais, 1049-001 Lisboa, Portugal}

\author{V. R. Vieira}
\address{Departamento de F\'{\i}sica and  CFIF, Instituto Superior 
T\'ecnico, Av. Rovisco Pais, 1049-001 Lisboa, Portugal}

\date{\today}

\maketitle

\begin{abstract}
We present a  mean-field study  of   superconductivity in a generalized
$N$-channel cubic Anderson lattice at $U=\infty$
taking into account the effect of a nearest-neighbor attraction $J$. The 
condition $U=\infty$ is implemented within the slave-boson formalism 
considering the slave bosons to be condensed.
We  consider the $f$-level occupancy  ranging from
the mixed valence regime to the Kondo limit
and study the dependence of the critical temperature 
on the various model parameters for each 
of three possible Cooper pairing symmetries (extended s, d-wave and p-wave
pairing) and find interesting crossovers.
It is found that the  $d-$ and $p-$ wave order parameters
have, in general, very similar critical temperatures. The extended s-wave
pairing seems to be relatively more stable  for electronic densities
per channel close to one and for large values of the superconducting interaction $J$.
\end{abstract}
\vspace{0.3cm}
\begin{multicols}{2}
\section{Introduction}

The superconducting behavior of heavy-fermion materials
has attracted much attention due to its non-conventional
properties. \cite{varma85}
Despite the large amount of work trying to understand heavy-fermion
superconductivity, the normal state properties, the symmetry
of the order parameter, the origin of superconductivity and
the interplay between superconductivity and magnetism are still
interesting and open questions. 

Some of these materials, 
such as  $UAgCu_4$, $UCu_7$, $U_2Zn_{17}$, 
order antiferromagnetically at low temperatures while
others (such as $UBe_{13}$, $CeCu_2Si_2$, $UPt_3$) 
order in a superconducting state and others show no ordering (such as
 $CeAl_3$, $UAuPt_4$, $CeCu_6$, $UAl_2$).  \cite{varma85}
There are materials 
which order both antiferromagnetically and become superconducting
as the temperature drops  ({\it e.g.} $URu_2Si_2$, $U_{0.97}Th_{0.03}Be_{13}$) 
and  it has recently been found that $UPd_2Al_3$ 
shows coexistence of superconductivity
and local moment antiferromagnetism. \cite{bernhoeft98}
 All  these materials have very
large specific heat coefficients $\gamma$, indicating very large effective
masses, hence the designation  {\it heavy fermions}.

The superconducting properties of a system depend on the type of ground state
that the system exhibits in the normal phase. 
The large specific heat 
$\gamma-$coefficient  can have 
two very different origins: 
a Kondo-impurity behavior, \cite{hewson} 
in which case $\gamma$ behaves as the inverse of the
Kondo temperature $T_K$; or a  Kondo-lattice behavior,
in which case $\gamma$ is controlled by a large density of states
at the Fermi energy. The large density of states  arises
from a hybridization mechanism between the conduction band and localized
electronic states ($f-$states, say). \cite{newns87} 

Even though the large effective masses indicate strong correlations between
the electrons, these behave in many cases as essentially ``free'', with 
renormalized parameters, as explained by the Fermi liquid theory. However,
 there has recently been  growing evidence that other materials have properties
that do not fit the Fermi liquid picture. \cite{jpcm}
The reason could be 
either  disorder, \cite{disord} vicinity to a quantum phase transition,
\cite{qpt} or  unusual
impurity-like behavior such as the one described by generalized models, as
the $n$-channel Kondo model. \cite{nchannel} The n-channel Kondo lattice
shows interesting behavior and it has been shown to be an incoherent metal
at low temperatures with a residual entropy that is usually lifted via
ordering at very low temperatures. \cite{ic}

A consistent description of the overall properties of the heavy-fermion
behavior has been achieved assuming that a generalization of the impurity
Anderson model to the lattice case is valid. \cite{newns87,Millis87}
In the Anderson lattice the energy of a single 
electron in an $f-$orbital (e. g.
$4f^1$)
is 
$\epsilon_0$, and the energy of two electrons 
in the same $f-$orbital ($4f^2$) is $2\epsilon_0+U$, where $U$ is 
the on-site
Coulomb repulsion. The energy of the  $4f^2$ state  is 
much larger than the energy of the $4f^1$ state. Moreover, these systems
are often characterized  by large angular momentum, due to the
spin-orbit coupling. \cite{hewson,newns87} In general, both the large values
of $U$ and the large total angular momentum  must be included
in any model used to describe the properties of
heavy-fermion materials.

The SU(N) Anderson lattice Hamiltonian is believed to give a good description
of the normal state of Kondo-lattice systems. \cite{newns87} 
The limit $U=\infty$  is considered in many
calculations since the experimental $U$ values are large.
The Anderson lattice model predicts Fermi-liquid like
properties in the normal non-magnetic state. The theoretical results give
a good 
description 
of many materials and explain the main features at
low temperatures such as universality, large effective masses, the Kondo
resonance at the Fermi level. At the single-impurity level the
picture is clear. In the Kondo limit the $f$-level has an occupation
close to one leading to a localized spin that is shielded by a conduction
electron spin cloud. This compensation of the spin explains why some
of these compounds do not order magnetically. The main point to be explained
in the lattice case
is the competition between the Kondo compensation of the localized spins
and the magnetic interactions between them. In these materials this interaction
is mediated by the conduction electrons ($RKKY$-type).
Actually, since the Kondo temperature is very
small it is difficult to explain why the RKKY does not always prevail. Related to
this competition is the effectiveness of the compensating cloud around each 
$f$-level. The size of this cloud has been subject of controversy. Arguments
show that it should be a large scale of the order of 
$v_F/T_K$ \cite{exhaust1}
but other arguments claim to be $\sim a$
($a$ is the lattice constant). \cite{exhaust2}
This is a relevant issue in the lattice case related to Nozi\`{e}res exhaustion
problem which states that there are not enough conduction electrons to screen
the $f$-levels.

To increase the complexity the system may also order into a superconducting
state. Many questions have been raised starting from the result that the 
discontinuity of the specific heat at $T_c$ is large, of the order of the
specific heat itself 
in the normal phase (which originates in the heavy fermions). This indicates
that pairing occurs between  the heavy $f$-level electrons, which will then 
form the  condensate. Within the Anderson lattice
model the strong correlations and the hybridization are responsible for
the high effective masses and  it has been proposed that the mechanism for
superconductivity lies in  the strong
Coulomb interaction between the $f$-electrons, not in a  phonon mediated attraction.

Using Coleman's \cite{coleman84} slave boson
formalism together with a large-$N$ approach, various attempts have been
made to search for the existence of superconducting
instabilities in the infinite-$U$ Anderson-lattice model. It was proposed
\cite{lavagna87} that  slave bosons fluctuations 
can provide an effective attraction
between the electrons to leading order in $1/N$.
Later, a calculation of the electron-electron 
scattering amplitude to order $1/N^2$
 revealed an effective
attractive interaction in the $p$ and $d$ channels, which was interpreted as
a manifestation of the $RKKY$ interaction, showing 
that spin fluctuations are an important mechanism.
\cite{houghton88} 

Assuming that the normal state is a Fermi liquid, 
several other studies of  superconductivity have been carried out
on the Anderson lattice model and  generalizations of it.
\cite{robaszkiewicz87,nathanson94,Gayatri94,romano97}
By adding an attractive nearest-neighbor interaction between the $f$-electrons,
so as to explicitly provide an attractive channel leading to superconductivity,
a mean-field study has been carried out as a function of the local repulsion
$U$.
Romano, Noce, and Micnas, \cite{romano97} have found a superconducting
ground state for finite values of $U$,
but no superconductivity was found for 
large values of onsite Coulomb repulsion, in 
the Anderson lattice. 
This is so because the authors consider 
the Kondo regime (this is, $\epsilon_0\ll \mu$\ where
$\mu$ is the chemical potential), 
where the occupation number of an $f-$orbital, $n_f$, 
is close to two for small $U$.
Therefore, upon increasing the interaction $U$, this number
is  reduced to one, blocking charge transport in the $f$-band. 

In this paper we carry out a mean field study
of superconductivity in the $U=\infty$ Anderson lattice where
an attractive interaction between neighboring $f$-orbitals is
explicitly introduced in order to simulate an 
effective interaction (which might have 
various causes) leading
to superconductivity. Since $U=\infty$, we  are restricted
to $f$-level occupancies in
the range $0<n_f<1$.
In the mixed valent regime, where $n_f$ is between zero and one, 
charge movement is allowed among the $f-$ orbitals, even 
when $U=\infty$. 
We study the dependence of the critical temperature 
and $f$-level ocupancy on the various model parameters for 
different Cooper pairing symmetries. 
The paper is organized as follows: in section \ref{s1}
we present the model Hamiltonian we use in our study and derive the mean field equations.
Particular attention is paid on the form of the superconducting pairing term.
In section \ref{s2} we present our calculations of the critical temperature
as function of the several parameters of the model and we 
summarize our findings in section \ref{s3}. 

\section{The Model Hamiltonian}
\label{s1}
We consider an extended version of the Anderson lattice 
model, 
which includes a density-density attraction between the electrons
occupying neighboring  $f$-orbitals. This form of interaction
enables us to consider three possible symmetries for electron pairing: $s$, $d$ and 
$p$-wave. The Hamiltonian is given by

\begin{equation}
H=H^0_c+H_f^0+H_{cf}+H_U+H_J\,,
\label{hi}	
\end{equation}
where
\begin{eqnarray}
H_f^0&=&\sum_{i,m}(\epsilon_0-\mu)f_{i,m}^{\dag}f_{i,m}\,,\\
H_c^0&=&\sum_{\vec k,m}(\epsilon_{\vec k}-\mu)
c_{\vec k,m}^{\dag}c_{\vec k,m}\,,\\
H_{cf}&=&V\sum_{i,m}\left(c_{i,m}^{\dag}f_{i,m}
+f_{i,m}^{\dag}c_{i,m}\right)
\label{hcf}\,,\\
H_U&=&U\sum_{i,m\neq m'}n_{i,m} n_{i,m'}\,,
\end{eqnarray}
and
\begin{equation}
H_J=\frac{1}{2}J\sum_{<i,j>,m,m'} n_{i,m} n_{j,m'}\,,
\end{equation}
where $i$ and $j$ are nearest neighbor sites and  $n_{i,m}=f_{i,m}^{\dag}f_{i,m}$.
The $c$ and $f$ operators  are fermionic and obey the usual anti-commutation
relations. The hybridization potential $V$ is assumed to be momentum
independent. The term $H_U$ represents the strong onsite repulsion 
between the $f$-orbitals and in the rest of this work 
we shall consider  $U=\infty$. The                 
term $H_J$ explicitly describes an effective attraction between 
neighboring $f$-sites  ($J<0$) which is responsible for  superconductivity. The
total angular momentum projection  $m$ takes on  $N$ values.
\cite{Millis87,coleman84}  We shall assume that the  local angular momentum
of the $f$-sites is half-integer and, therefore, that $N$ is even.

The term $H_J$  may be re-written
in momentum space as
\begin{equation}
H_J=\sum_{\vec{Q},\vec{k},\vec{k}'}\sum_{m,m'} 
\frac{J_{\vec{k},\vec{k}'}}{2}
{f^{\dagger}}_{\frac{\vec Q }{2}+\vec{k}',m} {f^{\dagger}}_{\frac{\vec Q}{2}-\vec{k}',m'}
f_{\frac{\vec Q}{2}-\vec{k},m'} f_{\frac{\vec Q}{2}+\vec{k},m} \,,
\label{hj}
\end{equation}
where the interaction  $J_{\vec{k},\vec{k}'}=J \sum_{\vec{\delta}} 
\exp{i(\vec{k}-\vec{k}')\cdot \vec{\delta}}$
and the summation over  $\vec{\delta}$ runs over the nearest neighbors. 
Considering  the case of a cubic
 lattice, the  interaction  $J_{\vec{k},\vec{k}'}$ may be  separated 
 into terms  with $s-$, $p-$ and $d-$wave  symmetries  as: \cite{bastide88}
\begin{eqnarray*}
J_{\vec{k},\vec{k}'}&=&J \left(
\eta^{(s)}_{\vec{k}}\eta^{(s)}_{\vec{k}'} +\sum_{i=x,y,z} 
\eta^{(p,i)}_{\vec{k}}\eta^{(p,i)}_{\vec{k}'}\right)\nonumber\\
&+&J \left(
\eta^{(d_{x^2-y^2})}_{\vec{k}}\eta^{(d_{x^2-y^2})}_{\vec{k}'}+
\eta^{(d_{r^2-3z^2})}_{\vec{k}}\eta^{(d_{r^2-3z^2})}_{\vec{k}'}\right)\,,
\end{eqnarray*}
where
\begin{eqnarray}
\eta_{\vec{k}}^{(s)}\  &=& \sqrt{\frac{2}{3}} \ \left[ \ 
\cos(k_x) + \cos(k_y) + \cos(k_z)\  \right]\,, \nonumber\\
\eta_{\vec{k}}^{(p,i)} &=& \sqrt{2} \ \sin(k_i)\,, \nonumber\\
\eta^{(d_{x^2-y^2})}_{\vec{k}} &=& \cos(k_x) - \cos(k_y) \,,\nonumber\\
\eta^{(d_{r^2-3z^2})}_{\vec{k}} &=& \frac{1}{\sqrt{3}}\left[\  
\cos(k_x) + \cos(k_y)-2\cos(k_z)\ \right]\,.
\label{simetrias}
\end{eqnarray} 
 Electron pairing in the superconducting phase will occur in the state
with total pair momentum  $\vec{Q}=0$.

We implement the condition $U=\infty$  within the slave-boson formulation due 
to Coleman,\cite{coleman84} in which the  empty $f$-site is represented
by a slave boson $b_i$ and  the physical operator $f_i$ in equation  (\ref{hcf})
is replaced with  ${b^{\dagger}}_i f_i$. Condensation of the slave-bosons can be described
by the replacement  $b_i \rightarrow <b_i>=<{b^{\dagger}}_i>=\sqrt{z}$. 
The mean-field treatment of the interaction term $H_J$ 
involves the usual decoupling of 
destruction and annihilation operators but,  in keeping with the spirit
of Coleman's slave boson formalism, we associate a boson operator with every
 $f$ operator in  (\ref{hj}) in order to   prevent 
double occupancy at the  $f$-sites.
Taking also  into account the boson condensation, we obtain  
the superconducting part of the mean-field 
Hamiltonian  from  the substitution:
  $f^{\dagger} f^{\dagger} f f \rightarrow z f^{\dagger} f^{\dagger} <z f f> + h.c.$.
Following these ideas we write down the effective  Hamiltonian as: 
\cite{note1}
\begin{eqnarray}
H_{eff} & = & \sum_{\vec k,m}\left( (\epsilon_{\vec k}-\mu)  c_{\vec k,m}^{\dag} c_{\vec k,m}
 + (\epsilon_f-\mu)f_{\vec{k},m}^{\dag}f_{\vec{k},m}\right) \nonumber\\
& + & \sqrt{z} V \sum_{\vec k,m}
(f_{\vec{k},m}^{\dag}c_{\vec k,m} + c_{\vec k,m}^{\dag}f_{\vec{k},m}) \nonumber\\
& + & \frac{1}{2}\sum_{\vec k,m} \left( z f_{\vec{k},m}^{\dag}f_{-\vec{k},m'}^{\dagger} \Delta_{\vec{k},m}+
z f_{-\vec{k},m'} f_{\vec{k},m} {\Delta^*}_{\vec{k},m}\right)\nonumber \\
& - & \frac{N_s}{2 J} \sum_m {\Delta^*}_m   \Delta_m \    +  
\ (\epsilon_f-
\epsilon_0) (z-1) N_s\,,
\label{heff}
\end{eqnarray}
where $N_s$ denotes the number of lattice sites and  
$\epsilon_f$ is the renormalized
energy of the $f$ orbitals due to the on-site repulsion.
 The angular momentum projection $m'=-m$ if electron 
pairing in   a singlet
state (s- or d-wave) is considered  and  $m'= m$ in the case of   p-wave pairing.
The gap function  $\Delta_{\vec{k},m} = \eta_{\vec k} \Delta_m$ and
the superconducting order parameter
$\Delta_m$ is given by
\begin{equation}
\Delta_m = \frac{z J}{N_s} \sum_{\vec k} 
\eta_{\vec k} <f_{-\vec{k},m} f_{\vec{k},m}>\,,
\label{eqgap}
\end{equation}
where $\eta_{\vec k}$ denotes  any of the possible pairing
symmetries considered in (\ref{simetrias}).

The  density of the  boson condensate $z$  minimizes
the free energy of the system and  $\epsilon_f$ is obtained 
after imposing
 local particle (boson+fermion) conservation at the  $f$-sites:
\begin{eqnarray}
z&=&1-\frac 1{N_s}\sum_{\vec k,m}<f_{\vec k,m}^{\dag}f_{\vec k,m}>\,,\label{z} \\
\epsilon_f-\epsilon_0 &=&-\frac V{2\sqrt{z}N_s}
\sum_{\vec k,m}\left(<f_{\vec k,m}^{\dag}c_{\vec k,m}>+
<c_{\vec k,m}^{\dag}f_{\vec k,m}>\right) \nonumber\\
&-& \frac{N_s}{z J}\sum_{m}   \Delta^*_m \,, \Delta_m
\label{etilf}
\end{eqnarray}
Equation (\ref{z}) states that the  mean number of electrons at an $f$-site is $1-z$.
 
In order to derive the gap equation and the spectrum of elementary excitations
we use the Gorkov Green's function approach. The  anomalous Green's functions 
that we need to consider are: 
\begin{eqnarray}
{\cal F}_{f,m}^{\dag}(\vec k,\tau-\tau')&=&
<T_{\tau}f^{\dag}_{\vec k,m}(\tau)
f^{\dag}_{-\vec k,-m}(\tau')>\,,\\
{\cal F}_{cf,m}^{\dag}(\vec k,\tau-\tau')&=&
<T_{\tau}c^{\dag}_{\vec k,m}(\tau)
f^{\dag}_{-\vec k,-m}(\tau')>\,,
\end{eqnarray}
and we  must also define three other Matsubara Green's functions: one 
that is associated with the conduction 
electrons, another one for the $f$-electrons
and the third one is related to the 
hybridization of the $f$ and $c$ bands:
\begin{eqnarray}
{\cal G}_{c,m}(\vec k,\tau-\tau') & = & -<T_{\tau}c_{\vec k,m}(\tau)
c^{\dag}_{\vec k,m}(\tau')>\,,\\
{\cal G}_{f,m}(\vec k,\tau-\tau') & = & -<T_{\tau}f_{\vec k,m}(\tau)
f^{\dag}_{\vec k,m}(\tau')>\,,\\
{\cal G}_{cf,m}(\vec k,\tau-\tau') & = & -<T_{\tau}c_{\vec k,m}(\tau)
f^{\dag}_{\vec k,m}(\tau')>\,,
\end{eqnarray}
After   Fourier  transforming  these functions into frequency space,
we  may write down  their equations of motion (Gorkov's equations) 
according to the Hamiltonian (\ref{heff}):
\begin{eqnarray}
(-i\omega_n&+&\epsilon_f-\mu) {\cal G}_{f,m}(\vec k,i\omega_n)
+V\sqrt{z}{\cal G}_{cf,m}(\vec k,i\omega_n)\nonumber\\
&+&Jz^2\Delta_m(\vec k)
{\cal F}_{f,m}^{\dag}(\vec k,-i\omega_n)=-1\,,
\label{al1}
\end{eqnarray}
\begin{equation}
(-i\omega_n+\epsilon_{\vec k}-\mu) {\cal G}_{cf,m}(\vec k,i\omega_n)
+V\sqrt{z}{\cal G}_{f,m}(\vec k,i\omega_n)=0\,,
\label{al2}
\end{equation}
\begin{equation}
(-i\omega_n-\epsilon_{\vec k}+\mu) {\cal F}_{cf,m}^{\dag}(\vec k,i\omega_n)
-V\sqrt{z}{\cal F}_{f,m}^{\dag}(\vec k,i\omega_n)=0\,,
\label{al3}
\end{equation}
\begin{eqnarray}
(-i\omega_n&-&\epsilon_f+\mu) {\cal F}_{f,m}^{\dag}(\vec k,i\omega_n)
-V\sqrt{z}{\cal F}_{cf,m}^{\dag}(\vec k,i\omega_n)\nonumber\\
\label{g2}
&+&Jz^2\Delta_m^{\dag}(\vec k)
{\cal G}_{f,m}(-\vec k,i\omega_n)=0\,. 
\label{al4}
\end{eqnarray}
Diagonalization of the above equations yields the energies of
the poles of the Green's functions (excitation energies) and
the corresponding residues (coherence factors).
The solutions are of the form
\begin{equation}
{\cal G}(\vec{k}, i\omega) = - \sum_{i=1,2} \sum_{\alpha=\pm} \frac{u_i^{\alpha}}{i
\omega_n + \alpha E_i}\,.
\end{equation}
The coherence factors, $u_i^{\alpha}$ and the excitation energies, $E_i$ are given
in the Appendix. 

The Green's functions have to be determined self-consistently
using the mean field equations (\ref{eqgap})-(\ref{etilf}). These equations 
can be rewritten in terms of Green's functions as
\begin{eqnarray}
z & = & 1- \frac{T}{N_s} \sum_{\vec{k},m} \sum_{i\omega_n} {\cal G}_{f,m} (\vec{k},
i\omega_n) \label{novaz}\,, \\
\epsilon_f - \epsilon_0 & = & - \frac{VT}{\sqrt{z}N_s} \sum_{\vec{k}, m} \sum_{i
\omega_n} {\cal G}_{cf,m} (\vec{k}, i\omega_n) \nonumber \\
& - & \frac{N_s}{zJ} \sum_{m} \Delta_m^{*}
\Delta_m \,,\\
\Delta_m & = & \frac{zJT}{N_s} \sum_{\vec{k}} \sum_{i\omega_n} \eta_{\vec{k}} 
{\cal F}_{f,m} (\vec{k}, i\omega_n) \,,
\label{novadelta}
\end{eqnarray}
For a given number of particles per site, $n$, these
equations  must be supplemented with the  particle conservation condition 
which yields the chemical potential $\mu$ for any temperature:
\begin{equation}
n =  1-z+\frac{T}{N_s} \sum_{\vec{k},m} \sum_{i\omega_n} {\cal G}_{c,m} (\vec{k},
i\omega_n)\,.
\label{mean}
\end{equation}

\section{Results}
\label{s2}
In what follows we consider a cubic lattice in  which the 
conduction band dispersion has the simple tight-binding form: 
$$
\epsilon_{\vec k}=-2t\sum_{i=x,y,z} \cos(k_i)\,,
$$
so that  $D=6t$ is half the bandwidth.
We have used the  subroutine hydrd.f from MINPACK in order to solve the 
four coupled equations (\ref{novaz})-(\ref{mean}).

The possible  pairing symmetries expressed in Eq. (\ref{simetrias}) 
have been studied separately. The two $\eta_{\vec{k}}$ functions corresponding 
to the $d$-wave symmetry
in equation (\ref{simetrias}) describe
different spatial orientations of the angular momentum of the Cooper pairs and 
give degenerate solutions. This same remark also applies to the three $p$-wave 
$\eta_{\vec{k}}$ functions in Eq. (\ref{simetrias}).

The critical temperatures $T_c$ are obtained solving the mean-field equations
using the normal state Green's functions. On the other hand, the study
of  $\Delta(T)$, $z(T)$, $\epsilon_f$(T), and the specific heat requires solving the 
mean-field equations with the full Green's functions.
In the normal phase, the slave boson condensation temperature, $T_z$, above which
$z=0$, is given by $T_z= (\epsilon_f-\mu)/\ln(N-1)$. If $N=2$, $z$ is always finite.
For larger values of $N$, and in particular in the limit $N\rightarrow \infty$,
$z\rightarrow 0$ as the temperature increases. Correspondingly, $n_f\rightarrow 1$
and the $f$-electron superconductivity is inhibited. Therefore, for large values of
$N$ it is expected that the mean-field theory will not yield superconductivity.
One then has to take into account the boson fluctuations. We will focus our
attention in the case $N=2$, relevant for instance for $Ce$ and $Yb$ materials, but we
will return to this point later.

\begin{figure}
\epsfxsize=8.0cm 
\epsfysize=8.0cm
\centerline{\epsffile{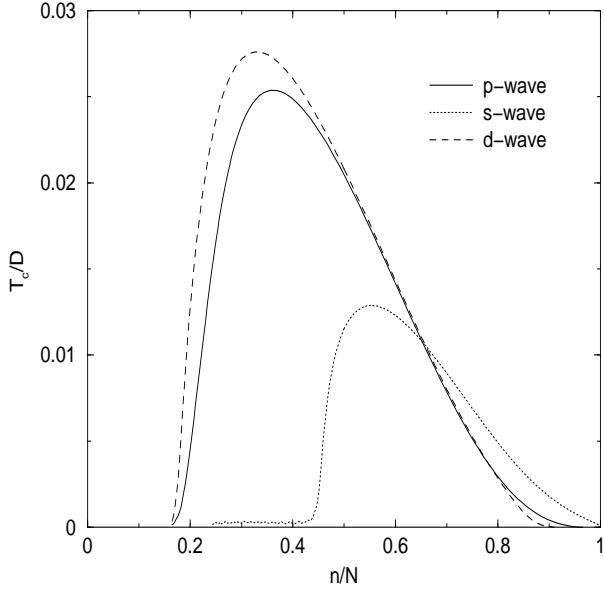}}
\caption{The critical temperature as function of the total density
of electrons per channel, for
the $U=\infty$ Anderson lattice. 
The parameters are $N=2$,  $\epsilon_f=-0.25D$,
$V=0.2D$, and $J=-0.5D$. The hopping integral $t=1$ and $D=6t$.} 
\label{tc_n_2_s_dw}
\end{figure}

Figure \ref{tc_n_2_s_dw}  shows
the behavior of the superconducting critical temperatures
$T_c$ as  function of the particle density  per channel $n/N$, for
each of the three pairing symmetries.  It is readily seen that the  critical temperatures 
associated with $d-$ and $p$-wave pairing follow similar trends and that the 
$d$-wave symmetry exhibits  the highest $T_c$ up to  densities of about $n/N\approx 0.6$.
At higher densities, a  crossover occurs into a regime where the extended $s$-wave 
pairing becomes the most stable, for the parameters considered.

The value of  $T_c$ vanishes at low densities  
because the $f$-level occupancy also becomes  small in that limit 
($z\rightarrow 1$)
and Cooper pairing occurs only between the 
$f$-electrons in the model under consideration. 
In the high density limit,
$T_c$ vanishes because  each $f-$level is almost fully occupied 
with one electron  ($z\rightarrow 0$), and 
freezing of the charge fluctuations 
[arising from the term $f^\dagger f^\dagger$
in (\ref{heff})] occurs because of  the infinite  on-site repulsion.

Heavy-fermion  behavior in the normal phase occurs
when the chemical potential $\mu$ lies close to
 the peak of the density of states (hence the strong effective mass).
 This peak is the equivalent of the 
 Kondo  resonance peak  which appears in the 
single-impurity problem. For the lattice problem, 
two strong peaks appear due to   hybridization between 
the conduction electron band and the dispersionless band of localized $f$-states, 
leading to the large  electron's effective mass. 
For densities above  $n/N\approx 0.7$ the chemical potential 
 becomes close to  the density of states peak  in the lower band. 

\begin{figure}
\epsfxsize=8.0cm 
\epsfysize=8.0cm
\centerline{\epsffile{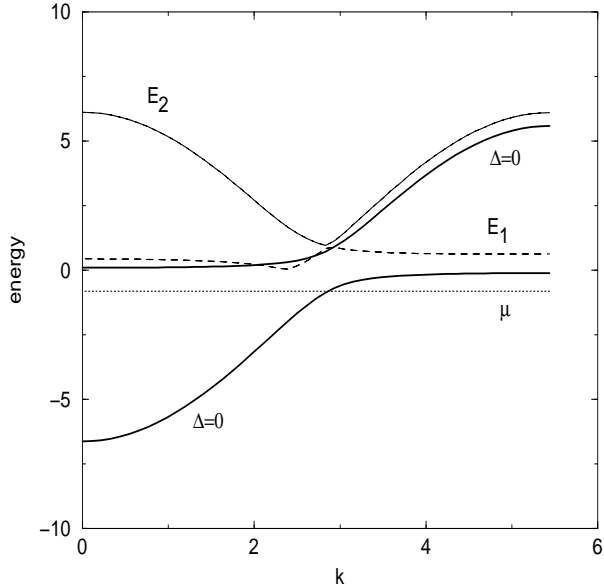}}
\caption{Band structure of the normal and
superconducting states, along the
direction $k_x=k_y=k_z$ in momentum space
($k=\sqrt{3}k_x$), at zero temperature. 
The total electronic density per
channel is $n/N=0.7$ and the other parameters are the same as in  Fig.
\protect{\ref{tc_n_2_s_dw}}.
 The symmetry of the superconducting
order parameter is extended $s-$wave. 
The excitation energies $E_1$ and $E_2$ are given by
Eqs. (\protect{\ref{e1}}) and (\protect{\ref{e2}}).} 
\label{bandfig}
\end{figure}

In the superconducting phase the full solution of Eqs. 
(\ref{novaz})-(\ref{mean}) yields
a renormalized excitation energy spectrum.  In Fig. (\ref{bandfig}) 
we show the band
structure for  $n/N=0.7$ in the normal and superconducting phases. It is clear
that for this density $\mu$ is in the flat region of the band in
the normal state.

\begin{figure}
\epsfxsize=8.0cm 
\epsfysize=8.0cm
\centerline{\epsffile{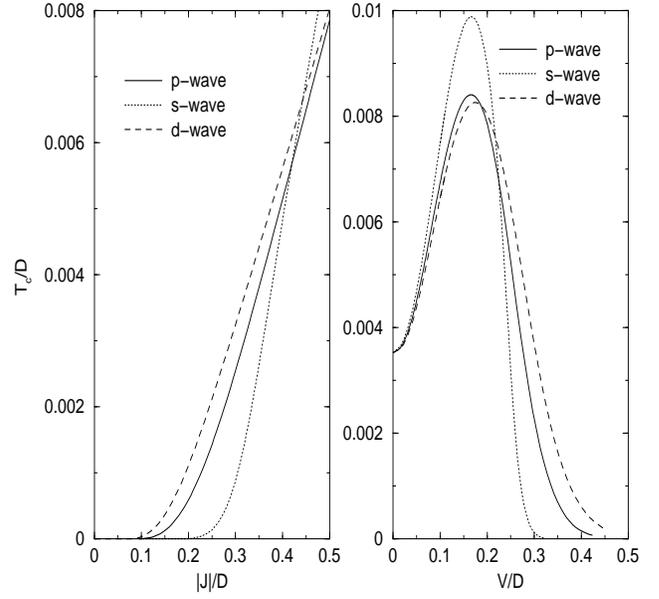}}
\caption{ Left panel:
The critical temperature $T_c$ as function of the coupling
$J$. Right panel:
The critical temperature $T_c$ as function of the hybridization
parameter $V$. The total electronic density per 
channel is $n/N=0.7$ and the other parameters are the same as in  Fig. 
\protect{\ref{tc_n_2_s_dw}}.}
\label{tc_J_n_2_s_dw}
\end{figure}

It is seen from Fig.
\ref{tc_n_2_s_dw}  that as the density per channel   $n/N$ approaches 1,
the value of   $T_c$ is strongly reduced until it eventually vanishes. From the
same figure one can also  see that  the critical temperature of the $s-$wave state 
at $n/N\approx 0.7$, for instance, is 
higher than that of the $d-$ or $p-$wave states. For the model parameters 
considered in  Fig. \ref{tc_n_2_s_dw} this means that
as   the temperature of a normal system is lowered, the  
system would first enter a superconducting state with extended s-wave symmetry.
On lowering further the temperature, the nature of the superconducting state 
becomes  a mixture of different symmetries. 
This
sequence of phase transitions would be
different had we chosen different parameters: our calculations show that 
if $J/D$ is less than about 0.4, then 
the critical superconducting temperature of a system with $n/N\approx 0.7$ 
would correspond to a $d$-wave order parameter (see 
left panel of Fig. \ref{tc_J_n_2_s_dw}). 

The dependence of $T_c$ on the parameters
$V$, $\epsilon_0$ and $J$ shows interesting crossovers.
If  $\epsilon_0$ is well below the chemical potential $\mu$ then 
the $f$-level is highly populated and 
the system  cannot become  superconducting unless the hybridization parameter 
$V$ is large enough.
On the other hand, if $\epsilon_0$ is not too low  a superconducting
ground-state is obtained even
for small values of
$V$. As  can be
seen from the right panel of Fig. \ref{tc_J_n_2_s_dw},
$T_c$ first increases with $V$
up to a maximum value, but as $V$ is further increased, 
large charge quantum fluctuations at the $f$ orbitals  are 
induced and superconductivity is destroyed.
Moreover, the $d-$ and $p-$wave 
superconductivity
seem to be more stable than the $s-$wave for large values of $V$. That this result 
is consistent with Fig. \ref{tc_n_2_s_dw}
can be easily understood as follows: upon increasing the hybridization between
the $f$-orbitals and the conduction band, the electron occupation in the 
$f$-sites is reduced and  Fig.
\ref{tc_n_2_s_dw} already showed that depletion of the $f$ band has the effect of
reducing $T_c$ and 
increasing
the stability of 
$d$-wave pairing relative to $p-$ and $s-$wave pairing.

\begin{figure}
\epsfxsize=8.0cm 
\epsfysize=8.0cm
\centerline{\epsffile{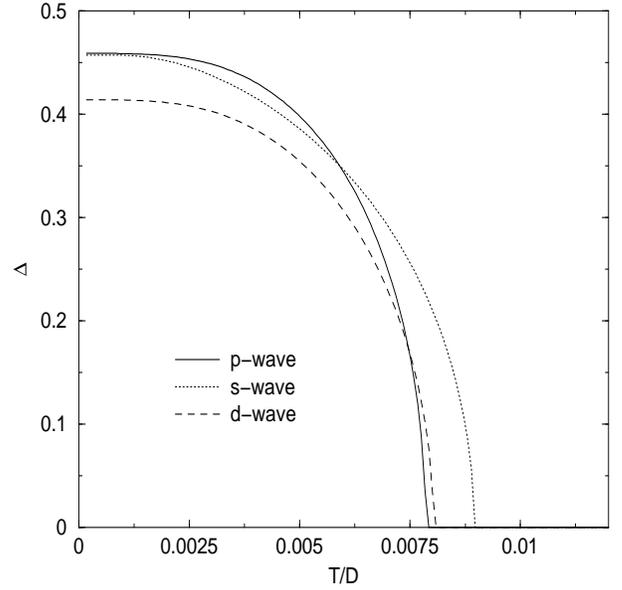}}
\caption{ Superconducting gap $\Delta(T)$
as functions of the temperature for
the three symmetries considered. The parameters are the same as in Fig.
\protect{\ref{tc_n_2_s_dw}}.}
\label{gapfig}
\end{figure}

The temperature dependence 
of the gap function in the superconducting phase is the 
standard one. In Fig. \ref{gapfig} we show a typical case.
The crossing of the $d-$ and $p-$order parameters
close to $T_c$ is related with the same crossing observed in $z(T)$.
Since close to $T_c$, $z(T)$ for the superconducting $d-$phase
becomes slightly higher than for the  superconducting $p-$phase,
the $d-$wave phase has an effective superconducting
coupling that is slightly higher than the $p-$wave coupling, 
leading to an higher $T_c$. 

In Fig. 5 we show the dependence of $T_c$ on the $f$-level
position. It is seen that the $d-$wave state
has always  a higher $T_c$ than the $p-$wave over  the range of
$\epsilon_0$ values considered. But the s-wave critical temperature 
exhibits a much stronger dependence on $\epsilon_0$. In particular, $s$-wave
pairing seems to be  more strongly depressed for low  $\epsilon_0$.

In a normal system at zero temperature the renormalized 
$f$-level energy $\epsilon_f$
is located above the chemical potential and 
$\epsilon_f - \mu$ is of the order
of the Kondo temperature for
 the equivalent single-imputity problem. Keeping the particle 
density fixed, both $\epsilon_f$ and  $\mu$
increase with temperature but the difference $\epsilon_f - \mu$ decreases. 
Our calculations show that $T_c$ is smaller than $\epsilon_f - \mu$ by a factor
of about 10 (see Fig \ref{tkfig}) over almost the entire range of 
densities considered in Fig. \ref{tc_n_2_s_dw}.
In Fig. \ref{tkfig} we present $T_c$, $\epsilon_f$, and $\mu$ for the
extended $s-$wave order parameter (the curves for the other symmetries
are qualitatively the same).
The susceptibility $dn_f/d\epsilon_f$,
in the region of densities characterized by $n/N\simeq 0.7$ or larger, 
is very small since the $f$-level density of states
is much larger than the $c$-level one, leading to a negative feedback changing the
chemical potential in such a way as to keep $\epsilon_f$ close to $\mu$.
This is very clear from  Fig. \ref{tkfig}, where $\epsilon_f$ is
indeed close to $\mu$, for electronic densities where the density of states
is large.
This is consistent with the picture that the pairing is developed by the
excitations of the system resulting from the Kondo compensated lattice.

Finally we calculate the specific heat for the various symmetries. 
The non-conventional
pairing symmetry leads to a power law behavior at 
low $T$ in the superconducting phase.
In Fig. \ref{heatfig} we show the specific heat for the
various symmetries as a function of temperature. The specific heat jump at the
transition is $\Delta C/C \sim 1.6, 1.3, 0.8$ for the $p,d,s-$symmetries
respectively. We have found that the specific heat at 
low $T$ has a  $T^2$ dependence for the $p,d-$ symmetries,
and has an exponential behavior for the $s-$wave case.

Considering now the effect of increasing the number of channels $N$,
we find that $T_c$ decreases by one order
of magnitude or more, as $N$ changes from $N=2$ to $N=4$.
For the parameters considered in the figures,
the effect is most dramatic for $s,p-$wave symmetries,  where superconductivity
is absent for $N\ge 4$. Furthermore, we found that the critical temperature
of a system with a  $d-$wave order parameter
is less sensitive to  the number of channels $N$, 
as  compared to the other symmetries.

\section{Summary}
\label{s3}
Heavy-fermions show a rich and complex behavior at low temperatures. In
particular, the interplay between magnetic correlations, the Kondo effect
and superconducting correlations is a difficult problem to solve. This is
further complicated since neither the mechanism nor the pairing symmetry
are fully established. In this paper we have focused on the superconducting
order assuming that the superconducting correlations are the dominant ones.
Using a generalized Anderson lattice model with nearest-neighbor attraction
between the $f$-electrons and with infinite-$U$ local Coulomb repulsion,
we studied the various pairing symmetries using a mean-field approach.
 In this way it is possible to compare the various
solutions in contrast to an approach where, starting from the normal
phase, the leading instabilities are identified.

The results show that there are several crossovers between the $s$, $d$ and $p$-wave
pairing symmetries as the parameters of the model are varied. In contrast
to a previous mean-field approach we find superconducting order, even though
$U=\infty$. The reason is that we focus on a regime where $0<n_f<1$, while the previous
work concentrated on a regime where $1<n_f<2$ (for finite $U$). Since we consider
only the case $U=\infty$, $n_f$ has to be smaller than one due to the Coulomb repulsion.
In the previous work as $U$ grows the density $n_f\rightarrow 1$ the $f$-electrons
become more localized inhibiting superconductivity. We find the same qualitative
behavior as we approach the Kondo regime from the mixed valent regime.
For small values of $\epsilon_0$ we tend to a regime where $n_f\rightarrow 1$ and
superconductivity is suppressed.

In the mean-field approach if $z\rightarrow 0$ ($n_f\rightarrow 1$) the gap function
$\Delta_m \rightarrow 0$. This happens for large densities $n/N\rightarrow 1$. 
For $n_f\rightarrow 0$ superconductivy is supressed, since the superconducting
coupling is among the $f-$electrons.
Also, if $N$ is large $z\rightarrow 0$ at
lower temperatures. In particular, $p$-wave and extended $s$-wave symmetries
are strongly suppressed. For $N=2$ $z$ is always finite. For larger values of $N$
in general it will be necessary to consider the boson fluctuations and
a treatment beyond mean-field will be required. For systems where the spin
degeneracy is low we expect the results to be qualitatively correct.

We have found that the $d$-wave and $p$-wave symmetries yield similar transition
temperatures. For
large nearest-neighbor attraction the extended $s$-wave pairing is preferred.
Otherwise, the $d$-wave symmetry seems to be more robust, in particular as $N$ grows.
Clearly,
we are not considering magnetic correlations in our mean-field study 
and therefore
the description applies to systems where there are no local moments (and therefore
$T_c<T_K$) and where $T_c>T_{RKKY}$.

We found that superconductivity is preferred in a mixed valent regime
(due to the infinite Coulomb repulsion). There are materials
that are mixed valent and superconductors \cite{mixed,wol,raus,ath,scmit,malik}.
 In the framework
of weak coupling BCS theory one would expect that the local magnetic character
of the $f$-states should be pair breaking. However, the heavy fermion
superconductivity in the Kondo limit (integer valent case) reveals that the
pairing is of another nature that compensates the pair breaking effects
of the local magnetic character. For materials such as $CeRu_3Si_2$ there
is a considerable mixed valent character and accordingly the effective masses
are not high. Also, the Wilson ratio is close to one indicating a conventional
weak-coupling BCS superconductor. Other mixed valent superconductors are
not conventional superconductors. It would be interesting to identify
systems that by changing the mixed-valent character could change the
superconducting temperature, $T_c$. In the framework of our model this would
require $f$-states with large $U$ values. The nearest-neighbor attraction
could be due to several mechanisms like spin fluctuations or slave boson
fluctuations (Coulombic nature).

\section{Acknowledgements}

We would like to thank P. Estrela for bringing Refs.\onlinecite{wol,raus,ath,scmit,malik} to
our attention. 
This research was supported by PRAXIS under grant number
2/2.1/FIS/302/94.

\appendix
\label{apa}
\section*{Poles and coherence functions for the Green's functions}
The algebraic solutions of Eqs. (\ref{al1}-\ref{al4}) for the Green's
functions ${\cal G}_{f,m}(\vec k,i\omega_n)$,
$ {\cal G}_{cf,m}(\vec k,i\omega_n)$, 
 $ {\cal F}_{f,m}^{\dag}(\vec k,i\omega_n)$, and 
$ {\cal G}_{c,m}(\vec k,i\omega_n)$ have the form 
\begin{equation}
{\cal G}(\vec k,i\omega_n)=-\sum_{i=1,2}\sum_{\alpha=\pm}
\frac{u_i^{\alpha}}{i\omega_n+\alpha E_i}\,,
\label{4poles}
\end{equation}
and the coherence factors  
$u_i^{\alpha}$ and the excitations energies $E_i$ are given below.

The energies $E_i$ have the form
\begin{eqnarray}
\label{e1}
E_1=\sqrt{\gamma/2-\sqrt{\gamma^2/4-\beta}}\,,\\
\label{e2}
E_2=\sqrt{\gamma/2+\sqrt{\gamma^2/4-\beta}}\,,
\end{eqnarray}
with $\gamma$ and $\beta$ given by
\begin{eqnarray}
\gamma=(\epsilon_f-\mu)^2+(\epsilon_{\vec k}-\mu)^2+2V^2z +\vert 
Jz^2\Delta(\vec k)\vert^2\,,\\
\beta=[(\epsilon_{\vec k}-\mu)(\epsilon_f-\mu)-V^2z]^2+\vert 
Jz^2\Delta(\vec k)\vert^2(\epsilon_{\vec k}-\mu)^2\,.
\end{eqnarray}
The $u_i^{\alpha}$ factors for ${\cal G}_{f,m}(\vec k,i\omega_n)$
are given by
\begin{eqnarray}
u_1^+= F(E_1+\epsilon_{\vec k}-\mu)X_1\,,\\
u_1^-= F(E_1-\epsilon_{\vec k}+\mu)Y_1\,,\\
u_2^+=-G(E_2+\epsilon_{\vec k}-\mu)X_2\,,\\
u_2^-=-G(E_2-\epsilon_{\vec k}+\mu)Y_2\,,
\end{eqnarray}
where the functions $X_i$ and $Y_i$ ($i=1,2$) are given by
\begin{eqnarray}
X_i=(\epsilon_{\vec k}-\mu)(\epsilon_f-\mu)&-&
(\epsilon_{\vec k}-\epsilon_f-2\mu)E_i\nonumber\\
&+&E_i^2-zV^2\,,\\
Y_i=(\epsilon_{\vec k}-\mu)(\epsilon_f-\mu)&+&
(\epsilon_{\vec k}-\epsilon_f-2\mu)E_i\nonumber\\
&+&E_i^2-zV^2\,,
\end{eqnarray}
and the functions $F$ and $G$ are given by
\begin{eqnarray}
F=\frac 1 {2E_1(E_2^2-E_1^2)}\,,\hspace{1cm}
G=\frac 1 {2E_2(E_2^2-E_1^2)}\,.
\end{eqnarray}
The $u_i^{\alpha}$ factors for ${\cal F}_{f,m}^{\dag}(\vec k,i\omega_n)$
are given by	
\begin{eqnarray}
u_1^+=-Jz^2\Delta(\vec k)
F(E^2_1-(\epsilon_{\vec k}-\mu)^2)\,,\hspace{.3cm}u_1^+=-u_1^-\,,\\
u_2^+=-Jz^2\Delta(\vec k)
G(E^2_2-(\epsilon_{\vec k}-\mu)^2)\,,\hspace{.3cm}u_2^+=-u_2^-\,.
\end{eqnarray}
The $u_i^{\alpha}$ factors for ${\cal G}_{cf,m}^{\dag}(\vec k,i\omega_n)$
are given by
\begin{eqnarray}
u_1^+=-V\sqrt z FX_1\,,\hspace{1cm}u_1^-=FY_1\,,\\
u_2^+=V\sqrt z GX_2\,,\hspace{1cm}u_2^-=-GY_2\,.
\end{eqnarray}
The $u_i^{\alpha}$ factors for ${\cal G}_{c,m}(\vec k,i\omega_n)$
are given by
\begin{eqnarray}
u_1^+=FQ_1\,,\hspace{1cm}u_1^-=-FR_1\,,\\
u_2^+=-GQ_2\,,\hspace{1cm}u_2^-=GR_2\,,
\end{eqnarray}
where
\begin{eqnarray}
Q_i=(\epsilon_{\vec k}-\mu-E_i)\vert Jz^2\Delta(\vec k)\vert^2
+(\epsilon_f+E_i)X_i\,,\\
R_i=(\epsilon_{\vec k}-\mu+E_i)\vert Jz^2\Delta(\vec k)\vert^2
-(-\epsilon_f+E_i)Y_i\,.
\end{eqnarray}		

\begin{figure}
\epsfxsize=8.0cm 
\epsfysize=8.0cm
\centerline{\epsffile{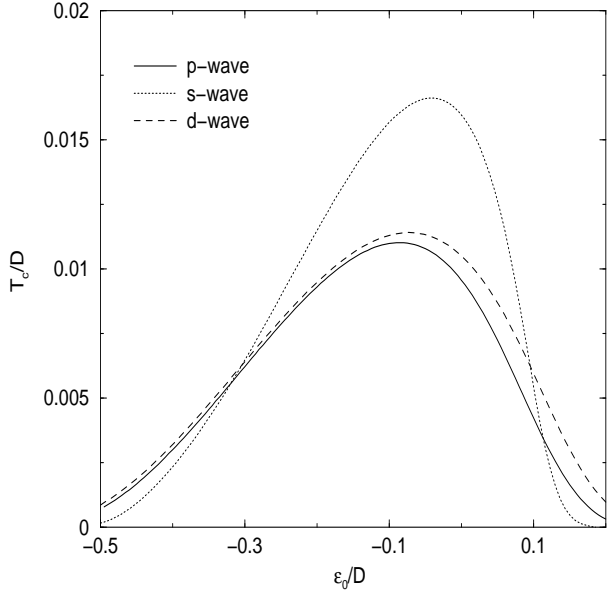}}
\caption{The critical temperature $T_c$ as function of
the $f$-level bare energy $\epsilon_0$.
The total electronic density per 
channel is $n/N=0.7$ and the other parameters are the same as in Fig. 
\protect{\ref{tc_n_2_s_dw}}.}
\label{tc_ef_s_dw}
\end{figure}

\begin{figure}
\epsfxsize=8.0cm 
\epsfysize=8.0cm
\centerline{\epsffile{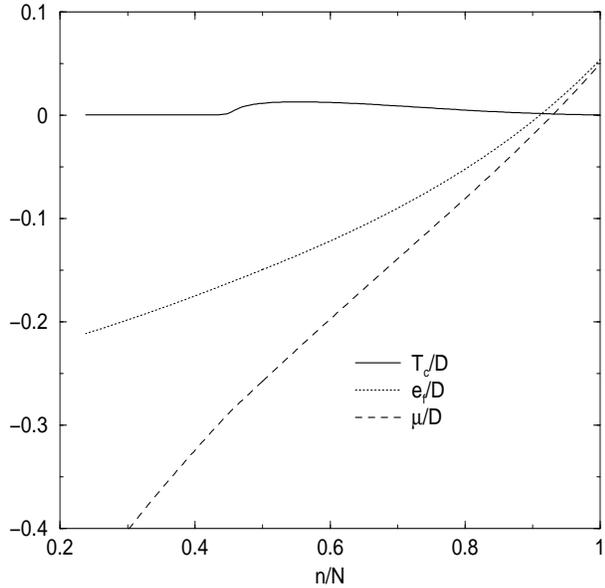}}
\caption{Critical temperature $T_c$, renormalized energy level
$\epsilon_f$, and chemical potencial $\mu$ as
function of $n/N$. The value of $\epsilon_f-\mu$
is much larger than $T_c$.
 The symmetry of the superconducting
order parameter is extended $s-$wave, and
the  other symmetries follow the same trends. 
The parameters are the same as in Fig.
\protect{\ref{tc_n_2_s_dw}}.}
\label{tkfig}
\end{figure}

\begin{figure}
\epsfxsize=8.0cm 
\epsfysize=8.0cm
\centerline{\epsffile{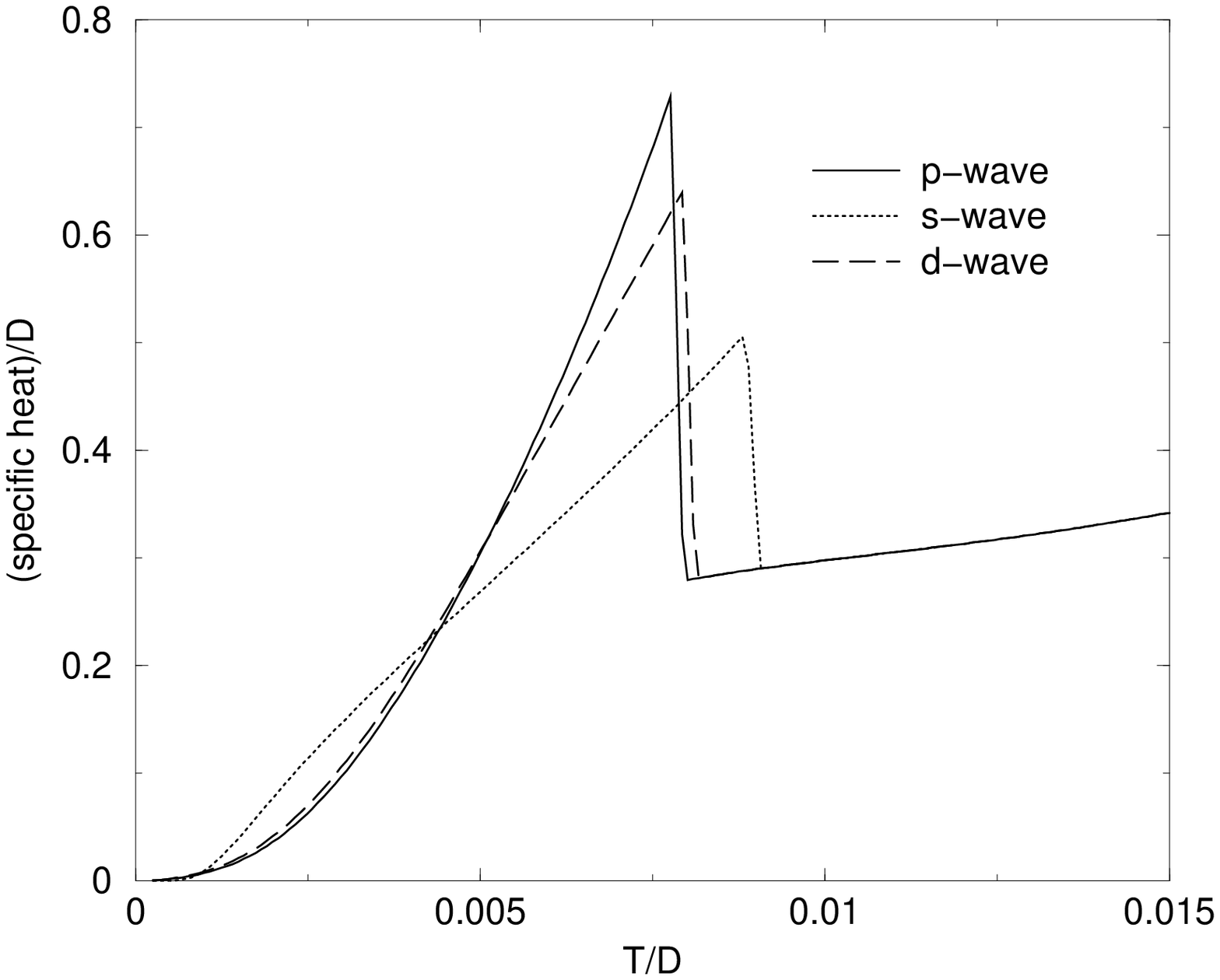}}
\caption{Specific heat $C(T)$ 
as function of the temperature for
the three symmetries considered. The parameters are the same as in Fig.
\protect{\ref{tc_n_2_s_dw}}.}
\label{heatfig}
\end{figure}

\end{multicols}

\begin{references}

\bibitem{varma85} C. M. Varma, Comments Solid  State Physics {\bf 11},
221 (1985); Z. Fisk et al., Science {\bf 239}, 33 (1988); P. Schlottmann,
Phys. Reports {\bf 181}, 1 (1989);
Robert H. Heffner
and Michael R. Norman, Comments Condensed Matter Physics {\bf 17},
361 (1996).

\bibitem{bernhoeft98} N. Bernhoeft, N. Sato, B. Roessli, N. Aso, A. Hiess,
G. H. Lander, Y. Endoh, and T. Komatsubara, 
Phys. Rev. Lett. {\bf 81}, 4244 (1998).

\bibitem{hewson} A. C. Hewson, {\it The Kondo problem to heavy-fermions},
(Cambridge, 1997).

\bibitem{newns87} D. M. Newns and N. Read, Adv. Phys. {\bf 36}, 799 (1987).

\bibitem{jpcm} See for instance the issue J. Phys. Cond. Matter {\bf 8}, 
number 48 (1986).

\bibitem{disord} V. Dobrosavljevic, T.R. Kirkpatrick and G. Kotliar, Phys.
Rev. Lett. {\bf 69}, 1113 (1992); O.O. Bernal et al., ibid, {\bf 75}, 2023
(1995).

\bibitem{qpt} B. Andraka and A.M. Tsvelik, Phys. Rev. Lett. {\bf 67}, 2886
(1991); A.M. Tsvelik and M. Reizer, Phys. Rev. B {\bf 48}, 9887 (1993).

\bibitem{nchannel} P. Nozi\`{e}res and A. Blandin, J. Phys. (Paris) {\bf 41},
193 (1980); P. Schlottmann and P.D. Sacramento, Adv. in Phys. {\bf 42}, 641
(1993); D.L. Cox and A. Zawadowski, Adv. in Phys. {\bf 47}, 599 (1998).

\bibitem{ic} M. Jarrell, H. Pang, D.L. Cox and K.H. Luk, Phys. Rev. Lett.
{\bf 77}, 1612 (1996); F.B. Anders, M. Jarrell and D.L. Cox, ibid. {\bf 78},
2000 (1997); M. Jarrell, H. Pang and D.L. Cox, ibid. {\bf 78}, 1996 (1997);
P.D. Sacramento and V.R. Vieira, Phys. Rev. B {\bf 58}, 11119 (1998).

\bibitem{Millis87} A. J. Millis and P. A. Lee, Phys. Rev B {\bf 35}, 3394
(1987).

\bibitem{exhaust1} P. Nozi\`{e}res, Ann. Phys. (Paris) {\bf 10}, 19 (1985);
V. Barzykin and I. Affleck, Phys. Rev. B {\bf 57}, 432 (1998).

\bibitem{exhaust2} J. Gan, J. Phys. Cond. Matter {\bf 6}, 4547 (1994);
V. Barzykin and I. Affleck, Phys. Rev. Lett. {\bf 76}, 4959 (1996).

\bibitem{coleman84} P. Coleman, Phys. Rev B {\bf 29}, 3035
(1984); Phys. Rev B {\bf 35}, 5072
(1987).

\bibitem{lavagna87} M. Lavagna, A. J. Millis, and P. A. Lee, Phys. Rev. Lett.
{\bf 58}, 266 (1987).

\bibitem{houghton88} A. Houghton, N. Read, and H. Won, Phys. Rev. B
{\bf 37}, 3782 (1988).

\bibitem{robaszkiewicz87} S. Robaszkiewicz, R. Micnas, and J. Ranninger,
Phys. Rev. B {\bf 36}, 180 (1987).

\bibitem{nathanson94} B. Nathanson and O. Entin-Wohlman,
Phys. Rev B {\bf 49}, 15377 (1994).

\bibitem{Gayatri94} Gayatri and P. Rudra, Phys. Rev. B {\bf 49},
14139 (1994).

\bibitem{romano97} A. Romano, C. Noce, and R. Micnas, 
Phys. Rev B {\bf 55}, 12640 (1997);
Acta Physica Polonica A {\bf 91}, 381 (1997).

\bibitem{bastide88} C. Bastide and C. Lacroix, J. Phys. C {\bf 21}, 3557
(1988).

\bibitem{note1} Note that $f_{i}^{\dagger}f_{i} = f_{i}^{\dagger} 
b_{i} f_{i} b_{i}^{\dagger}= f_{i}^{\dagger} f_{i} - 
f_{i}^{\dagger}f_{i} b_{i}^{\dagger}b_{i}$. The
second term will give zero when it acts on any state of the basis.
%
The anomalous term $f^{\dagger} f^{\dagger}$ however, violates the constraint
of no double occupancy and  therefore we take the order parameter as in eq. (10).
This same procedure has been used  in the context of the $t-J$ model in 
A. E. Ruckenstein, P. J. Hirschfeld
and J. Appel, Phys. Rev. B {\bf 36}, 857 (1987).
%


%
\bibitem{mixed} J. M. Lawrence, P.S. Riseborough and R.D. Parks, Rep. Prog. Phys. {\bf 44}, 1 (1981).
\bibitem{wol} D. Wohlleben and J. R\"ohler, J. Appl. Phys. {\bf 55}, 1904 (1984).
\bibitem{raus}U. Rauchschwalbe, W. Lieke, F. Steglich, C Godart,
 L.C. Gupta and R.D. Parks, Phys. Rev. B {\bf 30}, 444 (1984).
\bibitem{ath} K. S. Athreya, L.S. Hausermann-Berg, R.N. Shelton,
S.K. Malik, A.M. Umarji and  G.K. Shenoy , Phys. Lett. A {\bf 113}, 330 (1985).
\bibitem{scmit} W. Schmitt and G. G\"untherodt, J. Magn. Magn. Mater. {\bf 47}, 583 (1985).
\bibitem{malik} S. K. Malik, A.M. Umarji, G.K. Shenoy amd M.E. Reeves, J. Magn. Magn. Mater. {\bf 54}, 439 (1986).
%

\end{references}
\end{document}